\documentclass[doublespacing]{elsart}
\usepackage{amssymb}
\usepackage{amsmath}
\usepackage{amsfonts}
\usepackage{graphicx}
\usepackage{multicol}
\usepackage{wrapfig}

\begin{document}

\begin{frontmatter}
\title{Application of orthogonality principle to Endochronic and Mr\'{o}z models}
\author[a,b]{Nelly Point \corauthref{cor1}}
\ead{point@lami.enpc.fr}
\author[a]{, Silvano Erlicher}
\corauth[cor1]{Corresponding author. Tel: +33 1 64 15 37 40, Fax:
+33 1 64 15 37 41}
\address[a]{ Laboratoire Analyse des Mat\'{e}riaux
et Identification LAMI (ENPC/LCPC-Institut Navier),
6 et 8 av. B. Pascal, Cit\'{e} Descartes, Champs-sur-Marne, 77455
Marne-la-Vall\'{e}e, Cedex 2, France}
\address[b]{ Conservatoire National des Arts et M\'{e}tiers CNAM,
Sp\'{e}cialit\'{e} Math\'{e}matiques (442), 292 rue Saint-Martin,
75141 Paris, Cedex 03, France}
\begin{abstract}
A new description of Endochronic and Mr\'{o}z model is discussed. It
is based on the definition of a suitable pseudo-potential and the
use of generalized normality assumption. The key-point of this
formulation is the dependence of the pseudo-potentials on state
variables.
\end{abstract}
\begin{keyword}
Thermodynamics of solids \sep Endochronic theory \sep Generalized normality \sep Plasticity
\sep Pseudo-potentials \sep Mr\'{o}z model .
\end{keyword}
\end{frontmatter}


\section{Introduction}

A thermodynamically well-posed formulation of plasticity models
can be based on the definition of the Helmholtz free energy and of
the so-called pseudo-potential, from which the flow rules are
derived from the generalized normality assumption (orthogonality
principle) \cite{Ziegler58}, \cite{Ziegler87}, \cite{Fremond2002}.
It has been proven in \cite{Erlicher05} that the use of
pseudo-potentials with an additional dependence on \emph{state
variables} allows to describe classical plasticity models like
Prandtl-Reuss, Non-Linear Kinematic hardening models
\cite{Lemaitre90engl} as well as generalized plasticity
\cite{Lubliner93} and endochronic theory \cite{Valanis71}. In this
paper, the results concerning endochronic theory are recalled in
order to expose the proposed approach and then a new description
of the model of Mr\'{o}z \cite{Mroz67} is suggested.

\section{Thermodynamic framework}

Under the assumption of infinitesimal and isothermal
transformations, the second principle of thermodynamics states that
the \emph{intrinsic} or \emph{mechanical\ dissipation} $\Phi _{1}$
must be non-negative:
\begin{equation}
\Phi _{1}\left( t\right) :=\mbox{\boldmath $\sigma$}:
\mbox{\boldmath $\dot{\varepsilon}$}-\dot{\Psi}\geq 0
\label{ClausiusDuhem}
\end{equation}
$\mbox{\boldmath $\sigma$}$ is the Cauchy stress tensor (belonging
to the set $\mathbb{S}^{2}$ of symmetric and second-order tensors),
$\Psi\left(\mathbf{v}\right)$ is the Helmholtz free energy density,
function of $\mathbf{v=}\left( \mathbf{\ }\mbox{\boldmath
$\varepsilon$},\chi _{1},...,\chi _{n}\right) $, the vector
containing all the state variables, namely the total strain tensor
and the tensorial and/or scalar internal variables $\mathbf{\chi
}_{1},...,\mathbf{\chi }_{n}$, related to the non-elastic evolution.
The \emph{non-dissipative} thermodynamic forces are defined as:
\begin{equation}
\mbox{\boldmath $\sigma$}^{nd}:\mathbf{=}\frac{\partial \Psi
}{\partial \mbox{\boldmath $\varepsilon$}},\ \ \ \ \ \
\mbox{\boldmath $\tau$} _{i}^{nd}:=\frac{\partial \Psi }{\partial
\chi _{i}}  \label{fnondiss}
\end{equation}
Let $\mathbf{q}^{nd}=\left( \mbox{\boldmath $\sigma$}_{1}^{nd},
\mbox{\boldmath $\tau$}_{1}^{nd},...,\mbox{\boldmath
$\tau$}_{n}^{nd}\right) $ be the non-dissipative forces vector and
$\mathbf{\dot{v}}$ be the vector of the fluxes, belonging to a
vector space $\mathbb{V}$. Then, let us introduce the
\emph{dissipative }thermodynamic forces vector $\mathbf{q}
^{d}=\left( \mbox{\boldmath $\sigma$}^{d},\mbox{\boldmath $\tau$}
_{1}^{d},...,\mbox{\boldmath $\tau$}_{n}^{d}\right)$ belonging to
the dual space $\mathbb{V}^{\ast }$, with $\mbox{\boldmath
$\sigma$}^{d}:= \mbox{\boldmath $\sigma$}-\mbox{\boldmath
$\sigma$}^{nd}$ and $\mbox{\boldmath
$\tau$}_{i}^{d}:=-\mbox{\boldmath $\tau$}_{i}^{nd}.$ Hence, the
inequality (\ref{ClausiusDuhem}) can be written as follows
\begin{equation}
\Phi _{1}\left( t\right) =\mbox{\boldmath $\sigma$}: \mbox{\boldmath
$\dot{\varepsilon}$}-\mathbf{q}^{nd}\cdot
\mathbf{\dot{v}=q}^{d}\cdot \mathbf{\dot{v}}\geq 0
\label{2ndprinc1}
\end{equation}
where the symbol $\cdot $ indicates the scalar product of two objects having
the same structure \cite[pg. 428]{JirasekBazant02}. A classical manner to
ensure that (\ref{2ndprinc1}) is fulfilled, is to assume the existence of a
pseudo-potential $\phi \left( \mathbf{\dot{v}}\right) $ and to impose that $%
\mathbf{q}^{d}\in \partial \phi \left( \mathbf{\dot{v}}\right) $. In
this article, the pseudo-potential $\phi $ is allowed to vary with
the state variables $\mathbf{v}$. Hence, the corresponding
assumption called \emph{generalized normality}, or orthogonality
principle, reads
\begin{equation}
\mathbf{q}^{d}\in \left. \partial \phi \left( \mathbf{\dot{v}}^{\prime }%
\mathbf{;v}\right) \right\vert _{\mathbf{\dot{v}}^{\prime }=\mathbf{\dot{v}}}
\label{Normality}
\end{equation}
where $\partial $ indicates the sub-differential operator
\cite{Rockafellar69} with respect to the generic flow
$\mathbf{\dot{v}}^{\prime }$ and computed for
$\mathbf{\dot{v}}^{\prime }=\mathbf{\dot{v}}$, the actual value of
the flow. Defining the dual pseudo-potential $\phi ^{\ast}$ by the
Legendre-Fenchel transform $\phi ^{\ast }\left(
\mathbf{q}^{d^{\prime }};\mathbf{v}\right) =\sup_{\mathbf{\dot{v}\in
}\mathbb{V}}\left( \mathbf{q}^{d\prime }\cdot \mathbf{\dot{v}}-\phi
\left( \mathbf{\dot{v};v}\right) \right) ,$ the \emph{dual normality
condition} reads:
\begin{equation}
\mathbf{\dot{v}}\in \left. \partial \phi ^{\ast }\left( \mathbf{q}^{d\prime
};\mathbf{v}\right) \right\vert _{\mathbf{q}^{d\prime }=\mathbf{q}^{d}}
\label{NormalityDual}
\end{equation}
Plasticity is rate independent, hence $\phi $ is a positively
homogeneous function of order 1 with respect to the fluxes
$\mathbf{\dot{v}}^{\prime }$. Therefore, the dual pseudo-potential
becomes $\phi ^{\ast }=\mathbb{I}_{\mathbb{E}},$ the indicator
function of a closed convex set $\mathbb{E}$ depending on the
dissipative forces but also on the states variables. Moreover, the
dissipation reads: $\Phi _{1}=\mathbf{q}^{d}\cdot
\mathbf{\dot{v}=}\phi \left( \mathbf{\dot{v};v}\right) $.

\section{A new formulation of endochronic models}

\begin{figure}[tbp]
\begin{center}
\includegraphics[width=12.5cm]{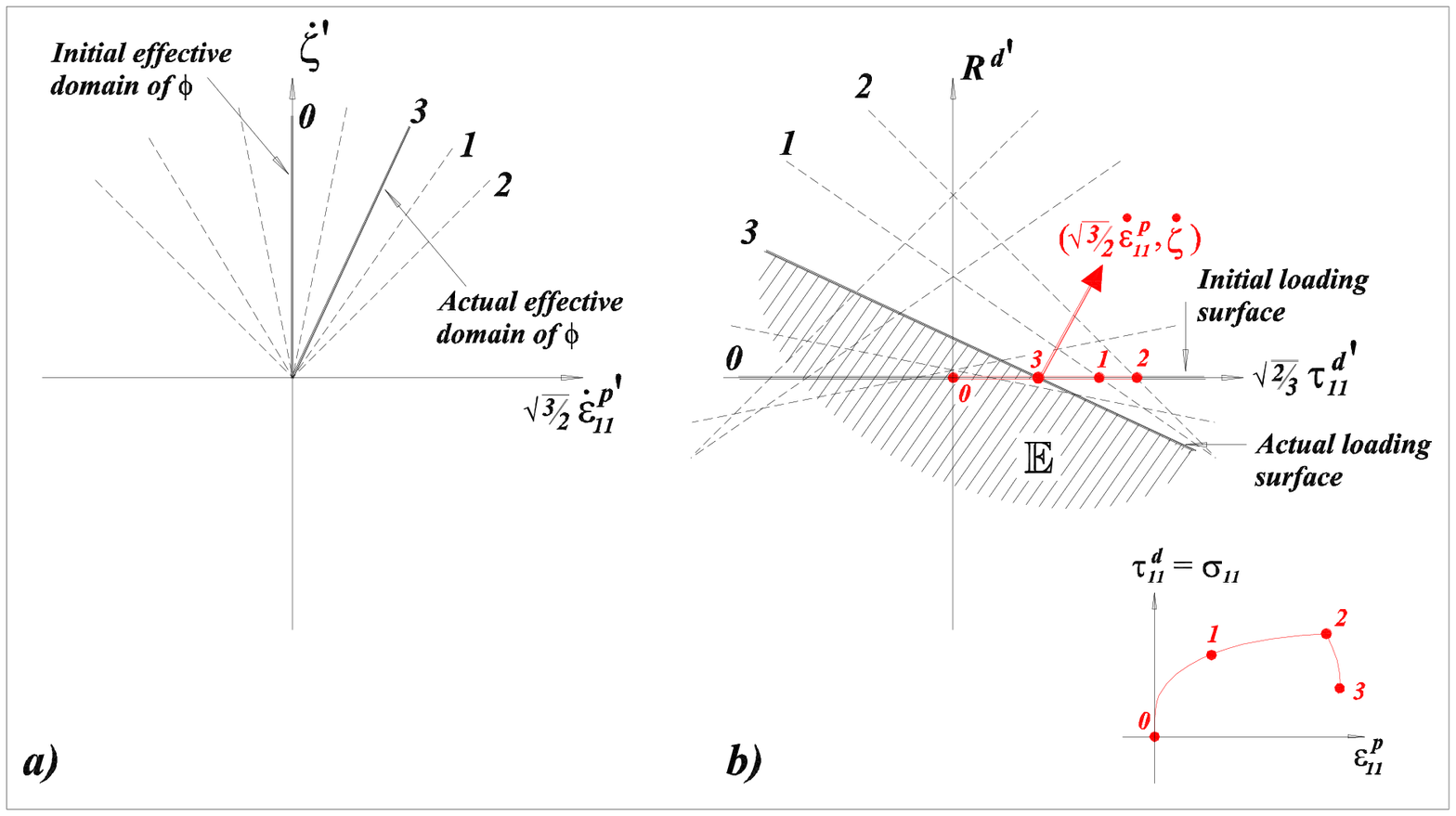}
\end{center}
\caption{Endochronic model. a) Several configurations of the domain $\mathbb{%
D}$. b) Configurations of the convex set $\mathbb{E}$ associated with $%
\mathbb{D}$. The actual state, $(\mbox{\boldmath $\tau$}^{d},R^{d})$, always
lies on the axis $R^{d^{\prime }}=0$, and the flux is normal to the boundary
of $\mathbb{E}$.}
\label{EndochronicDomains}
\end{figure}

Endochronic theory was first formulated by Valanis \cite{Valanis71}.
The model evolution is described by a convolution integral involving
the past values of $\mbox{\boldmath $\varepsilon$}$ and a positive
function $\mu$, the \emph{memory kernel}, depending on a positive
scalar variable called \emph{intrinsic time}. If $\mu $ is an
exponential, the integral expression can be rewritten as simple
differential equations. For an isotropic plastically incompressible
endochronic model, they read:
\begin{equation}
\left\{
\begin{array}{l}
tr\left( \mbox{\boldmath $\dot{\sigma}$}\right) =3K\ tr\left( %
\mbox{\boldmath $\dot{\varepsilon}$}\right)  \\
dev\left( \mbox{\boldmath $\dot{\sigma}$}\right) =2G\ dev\left( %
\mbox{\boldmath $\dot{\varepsilon}$}\right) \mathbf{-}\beta \ dev\left( %
\mbox{\boldmath $\sigma$}\right) \ \frac{\dot{\zeta}}{g\left( \zeta \right) }%
\end{array}%
\right.   \label{EndoFormulGen}
\end{equation}%
These relationships are equivalent to
\begin{equation*}
\mbox{\boldmath$\sigma$}\mathbf{=C:} \left( \mbox{\boldmath
$\varepsilon-\varepsilon$}^{p}\right) , tr\left(
\mbox{\boldmath$\dot{\varepsilon}$}^{p}\right) =0, \mbox{\boldmath
$\dot{\varepsilon}$}^{p}=\frac{dev\left(
\mbox{\footnotesize\boldmath $\sigma$}\right) }{2G/\beta }\
\frac{\dot{\zeta}}{g\left( \zeta \right) }.
\end{equation*}
with $\mathbf{C=}\left( K-\frac{2}{3}G\right) \mathbf{1\otimes
1+}2G\mathbf{I}$. The \emph{intrinsic time flow} is given by
$\dot{\vartheta}=\frac{\dot{ \zeta}}{g\left( \zeta \right)}$, where
$\zeta $ is the \emph{intrinsic time scale} and with $g(\zeta )\geq
0$ and $g(0)=1$. Generally $\dot{\zeta} =\left\Vert dev\left(
\mbox{\boldmath $\dot{\varepsilon}$}\right) \right\Vert $. \newline
The state variables and the associated non-dissipative thermodynamic
forces are represented by $\mathbf{v=}\left(
\mbox{\boldmath$\varepsilon$}, \mbox{\boldmath
$\varepsilon$}^{p},\zeta \right) $ and $\mathbf{q}^{nd}\mathbf{=}\left( %
\mbox{\boldmath $\sigma$}^{nd},\mbox{\boldmath $\tau$}^{nd},R^{nd}\right) $
respectively. The Helmholtz free energy $\Psi $ and the pseudo-potential are
chosen as follows :
\begin{equation}
\Psi \left( \mathbf{v}\right) =\frac{1}{2}\left(
\mbox{\boldmath
$\varepsilon-\varepsilon$}^{p}\right) :\mathbf{C}:\left(
\mbox{\boldmath
$\varepsilon-\varepsilon$}^{p}\right)   \label{PsiEndoIso}
\end{equation}%
\begin{equation}
\begin{array}{l}
\phi \left( \mathbf{\dot{v}}^{\prime };\mathbf{v}\right) =\frac{\left\Vert
dev\left[ \mathbf{C}:\left(
\mbox{\footnotesize\boldmath
$\varepsilon-\varepsilon$}^{p}\right) \right] \right\Vert ^{2}}{2Gg\left(
\zeta \right) /\beta }\dot{\zeta}^{\prime }+\mathbb{I}_{\mathbb{\bar{D}}%
\left( \mathbf{v}\right) }\left( \mathbf{\dot{v}}^{\prime }\right)  \\
\mathbb{\bar{D}}\left( \mathbf{v}\right) =\left\{
\begin{array}{l}
\left( \mathbf{\dot{v}}^{\prime }\right) \in \mathbb{V}\text{ \/ }\ tr\left( %
\mbox{\boldmath $\dot{\varepsilon}$}^{p^{\prime }}\right) =0,\text{ \ \ }%
\dot{\zeta}^{\prime }\geq 0, \\
\mbox{\boldmath $\dot{\varepsilon}$}^{p^{\prime }}=\frac{dev\left[ \mathbf{C}%
:\left( \mbox{\footnotesize\boldmath $\varepsilon-\varepsilon$}^{p}\right) %
\right] }{2Gg\left( \zeta \right) /\beta }\ \dot{\zeta}^{\prime }%
\end{array}%
\right\}
\end{array}
\label{fiEndoIso}
\end{equation}%
\noindent The condition $tr\left( \mbox{\boldmath
$\dot{\varepsilon}$}^{p^{\prime }}\right) =0$ introduces the plastic
incompressibility, while the third one characterizes the plastic
strain flow of endochronic theory. The positivity of
$\dot{\zeta}^{\prime }$ guarantees the positivity of $\phi $. When
the generic flux variable $\mathbf{\dot{v}}^{\prime }$ is equal to
the actual value of the flux $\mathbf{\dot{v}}$ , the first term of
$\phi $, in which the stress $\mbox{\boldmath
$\sigma$}^{nd}=\mathbf{C}:\left( \mbox{\boldmath
$\varepsilon-\varepsilon$}^{p}\right) $ is written as a function of
the state variables, is equal to the intrinsic dissipation $\Phi
_{1}$. To have zero viscous effect, the pseudo-potential is chosen
independent from $\mbox{\boldmath $\dot{\varepsilon}$}$. This
entails $\mbox{\boldmath $\sigma$}^{d}=0$. The dual pseudo-potential
$\phi ^{\ast }$ reads:
\begin{equation}
\phi ^{\ast }\left( \mathbf{q}^{d\prime };\mathbf{v}\right) =\mathbb{I}%
_{0}\left( \mbox{\boldmath $\sigma$}^{d^{\prime }}\right) +\mathbb{I}_{%
\mathbb{E}}\left( \mbox{\boldmath $\tau$}^{d^{\prime }},R^{d^{\prime }};%
\mathbf{v}\right)   \label{pseudoEndoIso}
\end{equation}%
with $\mathbb{E}=\left\{ \left( \mbox{\boldmath
$\tau$}^{d^{\prime }},R^{d^{\prime }}\right) \in \mathbb{S}^{2}\mathbb{%
\times R}\text{ / \ }f\leq 0\right\} $, and
\begin{equation}
\begin{array}{l}
f\left( \mbox{\boldmath $\tau$}^{d^{\prime }},R^{d^{\prime }};\mathbf{v}%
\right) =\frac{dev\left( {\footnotesize {\mbox{\boldmath $\tau$}^{d^{\prime
}}}}\right) :dev\left[ \mathbf{C}:\left( {\footnotesize {%
\mbox{\boldmath
$\varepsilon-\varepsilon$}}^{p}}\right) \right] }{2Gg\left( \zeta \right)
/\beta } \\
\text{ \ \ \ \ \ \ \ \ \ \ \ \ \ \ \ \ \ \ \ \ \ \ \ }-\frac{\left\Vert dev%
\left[ \mathbf{C}:\left( {\footnotesize {%
\mbox{\boldmath $ \varepsilon -\varepsilon
$}}^{p}}\right) \right] \right\Vert ^{2}}{2Gg\left( \zeta \right) /\beta }%
+R^{d^{\prime }}%
\end{array}
\label{fEndoIso}
\end{equation}%
The expression (\ref{fEndoIso}) defines the \emph{loading function of
endochronic models}. In Figure \ref{EndochronicDomains} the set $\mathbb{E}$
is represented in the $\left( \mbox{\boldmath $\tau $}^{d^{\prime
}},R^{d^{\prime }}\right) $ space together with the projection $\mathbb{D}$
of the effective domain $\mathbb{\bar{D}}$ when $g\left( \zeta \right) =1$.
As the system evolves, both sets change, due to their dependence on the
internal variables. At every instantaneous configurations, the set $\mathbb{D%
}$ is a straight line starting from the origin. The corresponding sets
\noindent $\mathbb{E}$ are half-planes orthogonal to $\mathbb{D}$.
Since $\Psi $ is
independent of $\zeta $ then $R^{nd}=-R^{d}=0$. Moreover, as $%
\mbox{\boldmath $\sigma$}^{d}=0$, at the actual stress state $%
\mbox{\boldmath
$\tau$}^{d}=\mathbf{C}:\left( {\footnotesize {%
\mbox{\boldmath
$\varepsilon-\varepsilon$}}^{p}}\right) $ and the loading function $f$ \emph{%
is always equal to zero}. In other words, $(\mbox{\boldmath $\tau$}%
^{d},R^{d})$ always belongs to the boundary of $\mathbb{E}$ , during
both loading and unloading phases, and \emph{all the states are
plastic states}. The dual normality conditions (\ref{NormalityDual})
lead to the endochronic flow rules:
\begin{equation}
\begin{array}{c}
\mbox{\footnotesize\boldmath $\dot{\varepsilon}$}^{p}=\frac{dev\left[
\mathbf{C}:\left( \mbox{\footnotesize\boldmath $\varepsilon
-\varepsilon$}^{p}\right) \right] }{2G\ g\left( \zeta \right) /\beta }\dot{%
\lambda},\ \ \dot{\zeta}=\dot{\lambda}\text{ ,\ }\dot{\lambda}\geq 0%
\end{array}
\label{flowendoIso}
\end{equation}%
Eqs. (\ref{pseudoEndoIso})-(\ref{fEndoIso}) and (\ref{flowendoIso}) prove
that endochronic models are associated in generalized sense. Moreover, since
$f$ is always equal to zero at the actual state, \emph{the consistency
condition is automatically fulfilled and cannot be used to compute }$\dot{%
\lambda}$. This situation, typical of endochronic theory, entails that the
plastic multiplier $\dot{\lambda}=\dot{\zeta}$ has to be defined by an
additional assumption. The standard choices are $g\left( \zeta \right) =1$
and $\dot{\vartheta}=\dot{\zeta}=\left\Vert dev\left(
\mbox{\boldmath
$\dot{\varepsilon}$}\right) \right\Vert $. More complex definitions can be
chosen \cite{Erlicher05}. It must be noticed that both flows $%
\mbox{\boldmath $\dot{\varepsilon}$}^{p}$ and $\dot{\zeta}$ can be different
from zero during unloading phases.

\section{New formulation of the Mr\'{o}z model}

\begin{figure}[tbp]
\begin{center}
\includegraphics[width=16.5cm]{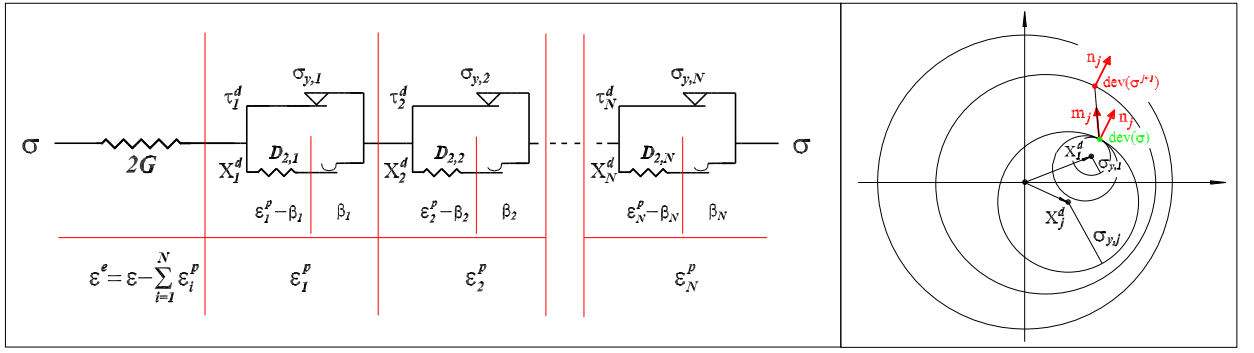}
\end{center}
\caption{Mr\'{o}z model. (a) Analogical scheme. (b) Loading
surfaces.} \label{Mrozl}
\end{figure}

Like for endochronic theory, the flow rules of the Mr\'{o}z model
\cite{Mroz67} can be deduced from a suited pseudo-potential using
the normality assumption. The Helmholtz free energy is defined as
follows:
\begin{equation*}
\begin{array}{l}
\Psi \left( \mathbf{v}\right) =\frac{1}{2}\left( \mbox{\boldmath
$\varepsilon$}-\sum_{i=1}^{N}\mbox{\boldmath
$\varepsilon$}_{i}^{p}\right) : \mathbf{C}:\left( \mbox{\boldmath
$\varepsilon$}-\sum_{i=1}^{N}
\mbox{\boldmath $\varepsilon$}_{i}^{p}\right)  \\
\text{ \ \ \ \ \ \ \ \ }+\frac{1}{2}\sum_{i=1}^{N}\left(
\mbox{\boldmath $\varepsilon$}_{i}^{p}-\mbox{\boldmath
$\beta$}_{i}\right) :\mathbf{D}_{i}:\left( \mbox{\boldmath
$\varepsilon$} _{i}^{p}-\mbox{\boldmath $\beta$}_{i}\right)
\end{array}
\end{equation*}
where $\mathbf{v=}\left( \mbox{\boldmath $\varepsilon$},
\mbox{\boldmath $\varepsilon$}_{1}^{p},...,\mbox{\boldmath
$\varepsilon$}_{N}^{p}, \mbox{\boldmath
$\beta$}_{1},...,\mbox{\boldmath $\beta$}_{N},\zeta _{1},...,\zeta
_{N}\right) $ are the state variables (total strain and, for each of
the N mechanisms introduced, plastic strain, kinematic and isotropic
hardening variables); $\mathbf{D}_{i}\mathbf{=} \left(
D_{1,i}-\frac{2}{3}D_{2,i}\right) \mathbf{1\otimes
1+}2D_{2,i}\mathbf{I}$ is the fourth-order tensor of the hardening
coefficients. The non-dissipative thermodynamic forces are deduced
\begin{equation}
\begin{array}{l}
\mbox{\boldmath $\sigma$}^{nd}=\frac{\partial \Psi }{\partial
\mbox{\boldmath $\varepsilon$}} =\mathbf{C}:\left( \mbox{\boldmath
$\varepsilon$}-\sum_{i=1}^{N}\mbox{\boldmath
$\varepsilon$}_{i}^{p}\right) =
\mbox{\boldmath $\sigma$} \\
\mbox{\boldmath $\tau$}_{i}^{nd}=\frac{\partial \Psi}{\partial
\mbox{\boldmath $\varepsilon$}_{i}^{p}}=-\mbox{\boldmath
$\sigma$}+\mathbf{X}
_{i}^{d}=-\mbox{\boldmath $\tau$}_{i}^{d} \\
\mathbf{X}_{i}^{nd}=\frac{\partial \Psi }{\partial \mbox{\boldmath
$\beta$} _{i}}=-\mathbf{D}_{i}:\left( \mbox{\boldmath
$\varepsilon$}_{i}^{p}-
\mbox{\boldmath $\beta$}_{i}\right) =-\mathbf{X}_{i}^{d} \\
R_{i}^{nd}=\frac{\partial \Psi }{\partial \zeta _{i}}=0=-R_{i}^{d}
\end{array}\label{nonDissForces1}
\end{equation}
Eqs. (\ref{nonDissForces1}) also report the relationships between
non-dissipative ($^{nd}$)\ and dissipative forces ($^{d}$). These
thermodynamic forces, together with the corresponding state
variables, define the analogical scheme depicted in Figure
\ref{Mrozl}a. The pseudo-potential $\phi $ is chosen as a sum of
$N$ pseudo-potentials, one for each of the $N$ mechanisms, i.e.
$\phi =\sum_{i=1}^{N}\left[ \phi
_{i}+\mathbb{I}_{\mathbb{\bar{D}}_{i}\left( \mathbf{v}\right)
}\right]$, with
\begin{equation*}
\begin{array}{l}
\phi _{i}=\left[ \sigma _{y,i}+\left( \mathbf{\varepsilon }
_{i}^{p}-\mathbf{\beta }_{i}\right):\mathbf{D}_{i}:\left(
\mathbf{n}_{j}\left( \mathbf{v} \right) -\mathbf{m}_{j}\left(
\mathbf{v}\right) \right) \right] \dot{\zeta} _{i}^{\prime }
\end{array}
\end{equation*}

\begin{equation*}
\mathbb{\bar{D}}_{i}\left( \mathbf{v}\right) =\left\{
\begin{array}{l}
\left( \mathbf{\dot{\varepsilon}}_{i}^{p^{\prime }},\dot{\zeta}_{i}^{\prime
},\mathbf{\dot{\beta}}_{i}^{\prime }\right) \text{\ / }tr\left( \mathbf{%
\dot{\varepsilon}}_{i}^{p^{\prime }}\right) =0,\text{ \ }\dot{\zeta}%
_{i}^{\prime }\geq \left\Vert \mathbf{\dot{\varepsilon}}_{i}^{p^{\prime
}}\right\Vert  \\
tr\left( \mathbf{\dot{\beta}}_{i}^{\prime }\right) =0\text{\ ,\ }\mathbf{%
\dot{\beta}}_{i}^{\prime }=\left( \mathbf{n}_{j}\left( \mathbf{v}\right) -%
\mathbf{m}_{j}\left( \mathbf{v}\right) \right) \dot{\zeta}_{i}^{\prime }%
\end{array}%
\right\}
\end{equation*}
Moreover (see also Figure \ref{Mrozl}b),
\begin{equation*}
\begin{array}{l}
\mathbf{n}_{j}\left( \mathbf{v}\right) =\frac{dev\left( \mbox{\boldmath $\tau$}%
_{j}^{d}\left( \mathbf{v}\right) \right) }{\left\Vert dev\left( %
\mbox{\boldmath $\tau$}_{j}^{d}\left( \mathbf{v}\right) \right)
\right\Vert}
\\
\mathbf{m}_{j}\left( \mathbf{v}\right) =\frac{dev\left(
\mbox{\boldmath $\sigma$}^{j+1}\left( \mathbf{v}\right)
-\mbox{\boldmath $\sigma$}\left( \mathbf{v}\right) \right)
}{\left\Vert dev\left( \mbox{\boldmath $\sigma$}^{j+1}\left(
\mathbf{v}\right) -\mbox{\boldmath
$\sigma$}\left( \mathbf{v}\right) \right) \right\Vert } \\
\mbox{\boldmath $\sigma$}^{j+1}\left( \mathbf{v}\right) =\mathbf{X}
_{j+1}^{d}\left( \mathbf{v}\right) +\frac{\mbox{\boldmath
$\sigma$}_{y,j+1}}{\sigma _{y,j}}dev\left( \mbox{\boldmath
$\sigma$}\left( \mathbf{v}\right) \mathbf{-X}_{j}^{d}\left( \mathbf{v}%
\right) \right)
\end{array}
\end{equation*}
where $\mathbf{m}_{j}$ is the vector defining Mr\'{o}z's translation
rule and $\mbox{\boldmath $\sigma$}^{j+1}$ corresponds to the\
target stress point, lying on the $(j+1)-th$ \emph{active} surface,
i.e. the largest one among those having the actual stress point on
its boundary; $\mathbf{n}_{j}$ is the normal to the active surface
at the actual stress point.

The formulation suggested here properly defines the duality between
the back-stresses $\mathbf{X}_{i}^{d}$ and the internal variables
$\mathbf{\beta }_{i}$. Moreover, it shows that the flows
$\mathbf{\dot{\beta}}_{i}$ may be different from zero only for
non-proportional loading, viz. $\mathbf{n}_{j}\neq \mathbf{m}_{j}$.
One can also notice that the term $\phi _{i}$ of the
pseudo-potential, contributing to the dissipation $\Phi _{1}$ when
$\dot{\zeta}_{i}^{\prime}=\dot{\zeta}_{i}\neq 0$, \emph{may be
negative} for non-proportional loading.

The dual pseudo-potentials can be computed by the Legendre-Fenchel
transform. Therefore :\\ $\phi^{\ast }=\sum_{i=1}^{N}\phi
_{i}^{\ast }=\sum_{i=1}^{N}\mathbb{I}_{\mathbb{E}_{i}\left(
\mathbf{v}\right) }$ with
\begin{equation*}
\mathbb{E}_{i}\left( \mathbf{v}\right) \mathbb{=}\left\{ \left( \mathbf{X}%
_{i}^{d^{\prime }},R_{i}^{d^{\prime }}\right) \text{ \ /\ }f_{i}\left(
\mathbf{X}_{i}^{d^{\prime }},R_{i}^{d^{\prime }};\mathbf{v}\right) \leq
0\right\}
\end{equation*}%
where the loading function $f_{i}$ is given by:

\begin{equation*}
\begin{array}{l}
f_{i}\left( \mathbf{X}_{i}^{d^{\prime }},R_{i}^{d^{\prime }};\mathbf{v}%
\right) =\left\Vert dev\left( \mathbf{\mbox{\boldmath $\tau$}}%
_{i}^{d^{\prime }}\right) \right\Vert +R_{i}^{d^{\prime }}-\sigma _{y,i} \\
+dev\left( \mathbf{X}_{i}^{d^{\prime }}\right) :\left( \mathbf{n}_{j}-%
\mathbf{m}_{j}\right) -\left( \mathbf{\mbox{\boldmath
$\varepsilon$}}_{i}^{p}-\mathbf{\mbox{\boldmath $\beta$}}_{i}\right) :%
\mathbf{D}_{i}:\left( \mathbf{n}_{j}-\mathbf{m}_{j}\right)
\end{array}%
\end{equation*}
At the actual state, one has $f_{i}= \left\Vert dev\left(
\mathbf{\mbox{\boldmath $\tau$}}_{i}^{d}\right) \right\Vert -\sigma
_{y,i}$, which is the usual definition for the loading function. The
normality conditions (\ref{NormalityDual}) lead to the well-known
Mr\'{o}z flow rules:
\begin{equation}
\mathbf{\dot{\varepsilon}}_{i}^{p}=\dot{\lambda}_{i}\mathbf{n}_{j},\text{ \
\ \ \ }\dot{\zeta}_{i}=\dot{\lambda}_{i},\text{ \ \ \ }\mathbf{\dot{\beta}}%
_{i}=\dot{\lambda}_{i}\left( \mathbf{n}_{j}-\mathbf{m}_{j}\right)
\label{KT*}
\end{equation}%
with the Kuhn-Tucker conditions $\dot{\lambda}_{i}f_{i}=0,$ $\dot{\lambda}%
_{i}\geq 0,$\ $f_{i}\leq 0$; recall that $f_{i}=0$ for all $i\leq
j$. The consistency conditions, viz. $\dot{f}_{i}\left(
\mathbf{X}_{i}^{d^{\prime }},R_{i}^{d^{\prime }};\mathbf{v}\right)
=0$ for $i\leq j$, lead to the computation of $\dot{\lambda}_{i}$
and then of the plastic flow
$\mathbf{\dot{\varepsilon}}^{p}:=\left(\sum_{i=1}^{j}
\dot{\lambda}_{i}\right)\mathbf{n}_{j}$.

\section{Conclusions}

Pseudo-potentials depending on states variables and the normality
assumption have been used to formulate the endochronic theory of
plasticity and the Mr\'{o}z model. This description helps to
investigate the thermodynamic properties of both models and makes
possible an insightful comparison with other classical and
non-classical plasticity theories.

\end{document}